\documentclass[epj,refree,nopacs,draft]{svjour}
\def\D{\Delta}\def\T{\theta}\def\a{\alpha}\def\I{M_I}\def\U{M_U}
\def\Z{M_Z}\def\ov{\overline}\def\un{\underline}\def\op{\oplus}\def\t{\times}
\def\ep{\epsilon}\def\b{\begin{equation}}\def\e{\end{equation}}
\def\be{\begin{eqnarray}}\def\ee{\end{eqnarray}}\def\ba{\begin{array}}
\def\ea{\end{array}}\def\rrep{{\un {126}}\oplus{\overline {\un {126}}}}
\begin{document}
\title{High scale perturbative gauge coupling in R-parity conserving SUSY 
$SO(10)$ with longer proton lifetime}  
\author{M.K. Parida \thanks{\emph{e-mail:} mparida@sancharnet.in}
\and B.D. Cajee}\institute{Physics Department, North-Eastern Hill University, 
Shillong 793022, India}
\date{Received: 4 July 2005}
\abstract{It is well known that in single step breaking of R-parity conserving SUSY $SO(10)$
that needs the Higgs representations $\rrep$  the  GUT-gauge coupling 
 violates  perturbative constraint at mass scales few times 
larger than the GUT scale. Therefore, 
if the $SO(10)$ gauge coupling is to remain perturbative  up to the Planck 
scale($\equiv 2\times 10^{18}$ GeV), the  scale $\U$ of the GUT symmetry 
breaking is to be bounded from below. The bound depends upon specific Higgs representations used for $SO(10)$ symmetry breaking but, as we find, can not 
be lower than $1.5\times 10^{17}$ GeV. In order to obtain such a high unification scale we propose a two-step $SO(10)$ breaking through 
$SU(2)_L\times SU(2)_R\times U(1)_{B-L}\times SU(3)_C$($g_{2L}\neq g_{2R}$) 
intermediate gauge symmetry.
We estimate potential   threshold and gravitational 
corrections to the running of gauge couplings and show that they can make the picture of perturbative GUT-gauge coupling running consistent at least up 
to the Planck scale.
We also show that when $SO(10) \to G_{2213}$ by Higgs representations 
$\un {210} \op \un {54}$, gravitational corrections alone with  negligible
threshold effects may
 guarantee such perturbative ~gauge ~coupling. 
The lifetime of the proton is found to increase by nearly $6$ orders over the present experimental limit for $p\rightarrow e^+\pi^0$.
For the proton decay mediated by ${\rm dim}.5$ operator a wide range of 
lifetimes is possible extending from the current experimental limit up to 
values $2-3$ orders longer.} 
\maketitle
\section{Introduction}\label{sec1}
~~~~In spite of its astounding  success the nonsupersymmetric 
standard model(SM) suffers from the well known gauge hierarchy problem.
 It fails to explain the available data on neutrino masses and mixings and also
fails to exhibit unification of the three known gauge couplings at  higher 
scales. One compelling reason to solve the gauge hierarchy problem is to go 
beyond the SM through weak-scale SUSY 
as in the minimal supersymmetric standard model(MSSM) \cite{1}. The MSSM has the added virtues that, in addition to explaining the origin of electroweak symmetry breaking, it provides a candidate for dark matter of the Universe.~If the 
SM fermion representations are extended by the addition of one
right-handed neutrino per generation and the corresponding extension is made
in  MSSM, the model can  account for
neutrino masses and mixings through seesaw mechanisms \cite{2,3,4}.

  Another  amazing aspect of  MSSM  has been noted to be  the unification  
of the three gauge couplings of disparate strengths and origins   when extrapolated to  as high a scale as $M_U=2\times 10^{16}$ GeV \cite{5}. However, the 
meeting of the three gauge coplings can be truly termed  the grand unification 
~\cite{6,7} of the three basic forces of nature provided the merged 
coupling constants evolve as a single gauge coupling at  higher
scales and  some simple ansatz for this have  been hypothesized through SUSY 
GUTs such as
SU(5), SO(10), $E_6$, and a number of others \cite{7,8,9}. 

 While R-parity violation as an automatic  consequence of MSSM spoils the 
predictive power of supersymmetric theories, an additional elegant feature of 
SUSY SO(10) breaking down to MSSM
is its potentiality to conserve R-parity. As the minimal left-right symmetric GUT  SO(10) contains the maximal subgroup 
$SU(2)_L\times SU(2)_R \times SU(4)_C$ of 
Pati-Salam \cite{6} 
which~in turn contains $SU(2)_L\times U(1)_R \times SU(4)_C$,~$SU(2)_L\times SU(2)_R\times U(1)_{B-L}\times SU(3)_C$ ($\equiv G_{2213}$),
$SU(2)_L\times U(1)_R\times U(1)_{B-L}\times SU(3)_C$ and 
$SU(2)_L\times U(1)_Y\times SU(3)_C$($\equiv G_{213}\equiv SM$) as its 
subgroups \cite{10}. Thus,
subject to the cosistency with the renormalization group constraints, SUSY 
$SO(10)$ gauge symmetry may break to the $SM$ gauge group directly in one-step or through an intermediate gauge 
symmetry to the MSSM \cite{11,20}. In addition to other superheavy representations needed to implement the GUT symmetry breaking,
two different popular choices of Higgs representations being extensively used
to obtain the  MSSM from  SUSY SO(10)
are $\un {16} \op {\ov{\un {16}}}$   and $\rrep $ . 
While the first choice violates R-parity, the second conserves it.
The Higgs representations  $\rrep$ in SUSY SO(10) have been found to solve 
a number of problems on fermion masses through renormalizable interactions. 
To cite a few, it rectifies the bad 
SU(5)-mass relation in the right direction in SO(10) to yield ${\rm m}_\mu=3{\rm m}_s$ . It attributes large atmospheric neutrino  mixing to 
$b-\tau$ ~unification and accommodates the masses and mixings of three 
neutrino flavors, in addition to the observed masses  and mixings of all 
other fermions,  via Type II seesaw 
mechanism \cite{12,13,14,15,21}.  
 However, because 
of large contribution to the $\beta$-function coefficient of the gauge 
coupling evolution, the presence of $\rrep$ in R-parity
coserving SUSY SO(10), in addition to other Higgs representations, violates 
perturbative constraint on the GUT gauge coupling($\alpha_G < 1$) even at mass scales few times larger than the GUT scale($=2\times 10^{16}$ GeV). Although
there might be deeper reasons to believe that the R-conserving SUSY at such scales could be nonperturbative, it is desirable to have a perturbative
theory at least up to the compactification  scale($M_{CS}\simeq 10^{17}$ GeV) or the Planck
scale ($M_{Pl}=2\times 10^{18}$ GeV).

~~~Proton decay is a necessary prediction of a number of  GUTs including SU(5) 
and SO(10). The decay mode $p\to e^+ \pi^0$ common to both SUSY and nonSUSY 
GUTs is mediated by superheavy gauge bosons carrying fractional charges and
the corresponding effective Lagrangian has a ${\rm dim}.6$ operator. In SUSY 
GUTs superpartners of fermions and heavy colour triplets of Higgs bosons 
give rise to new decay modes such as $p \to K^+ {\overline {\nu}}_{\mu}$,      $p \to K^+ {\overline {\nu}}_{\tau}$, and others. The mediation of 
the heavy superpartner  leads to a 
${\dim}.5$ operator in the effective Lagrangian for these supersymmetric decay 
modes. Recent experimental measurements provide improved limits on the lifetimes for both these types of decay modes,      
\be\tau \left(p\rightarrow e^+\pi^0\right) \ge 4\times 10^{33} \rm {years}
\label{eq1}\ee 
\be\tau(p \rightarrow K^+\overline{\nu}_{\tau})\ge 2.2\times 10^{33} \rm years \label{eq2}\ee \\
While eq.(1)  gives the bound $M_U \ge 5.6 \times 10^{15}$ GeV, eq.(2) yields
the limit on the superheavy colour triplet Higgssino mass as $M_{T_{\tilde C}} \ge  10^{17}$
 GeV. Although this has been treated as a severe constraint on SUSY
$SU(5)$ \cite{16}, easier methods  have been suggested to evade 
it \cite{17,18}.
In R-parity coserving SUSY SO(10) another interesting suggestions have been made to 
increase proton lifetime of supersymmetric decay mode  through
specific Yukawa textures, but in this case the GUT gauge couling
remains perturbative only upto $\mu={\rm few}\times 2\times 10^{16}$ GeV 
\cite{19}.     .     

 In this paper we show that with the similar choices of Higgs representations
as in the single step breakings of R-parity conserving SUSY SO(10), when the GUT gauge symmetry is allowed to break down to MSSM through $G_{2213}$-intermediate gauge symmetry investigated recently \cite{20}, perturbative GUT gauge coupling is ensured at least upto 
the  Planck scale due to threshold and gravitational corrections.
Although in this paper we have addressed the issue of perturbative gauge coupling up to the reduced Planck scale($=2\times 10^{18}$ GeV), we have checked that our method also works 
even if we use the Planck scale as $M_{Pl}\simeq 1.2 \times 10^{19}$ GeV
according to  the definition of the Particle Data Group. 
The realization  of perturbative grand unification in R-parity conserving
SO(10)  which has not been possible otherwise  is demonstrated for the first time in this paper. Other new contributions of the present paper compared to 
\cite{20} are derivations of gravitational corrections in the presence of~Higgs representations $\un {54}$ and $\un {210}\op \un {54}$ which contribute to 
SO(10) breaking near the GUT scale. Combining perturbative criteria with 
R-parity
conservation in SUSY SO(10) we obtain  lower bounds on the unification scale
in different cases.
Very significant increase of proton lifetimes is obtained  leading to  the greater stability of the particle.

  In Sec.2, we discuss the origin of high-scale violation of perturbation theory in SUSY SO(10). In Sec.3 we discuss analytically 
threshold and gravitational corrections.  In Sec.4 we show how 
these corrections elevate the unification scale so as to satify perturbative constraint on the GUT gauge coupling at least up to the Planck scale.
In Sec.5 we discuss  inrease in proton lifetimes in different cases. 
 Summary and conclusions are stated in Sec.6.\\

\section {Perturbative constraint and lower bounds on                          unification scale} 
\label{sec2}

 With R-parity conservation a minimal SO(10) model having  26 parameters has 
been identified to be the one with Higgs representations: $\un {210} \op\rrep\op \un {10}$ \cite{31} for which a very interesting method of proton lifetime increase  has  been 
suggested \cite{19}. In order to account for neutrino masses and mixings in SUSY SO(10) through Type II seesaw dominance, the realistic symmetry breaking pattern has been shown to require 
$\un {210}\op \un {54}\op\rrep\op \un {10}$ \cite{21} where both $\un {210}$ 
and $\un {54}$ are present. We will show that in this case with $G_{2213}$
intermediate breaking 
 gravitational corrections alone may be sufficient to guarantee perturbative
gauge coupling at higher scales.
But, in the single step breaking scenario above the GUT scale, not only these two  models but also other variants of R-parity conserving SUSY SO(10)  violate perturbation theory even at mass scales   $\mu={\rm few}\times 2\times 10^{16}$ GeV   
whenever the Higgs representations $\rrep$ are present in the model.
 
Above the GUT scale $(\mu> M_U)$ the GUT fine structure constant 
$\alpha_G(\mu) =\frac{g^2_G(\mu)}{4\pi}$, where $g_G=$GUT coupling, evolves at  one-loop level as \\

\be\frac{1}{\alpha_G(\mu)}=\frac{1}{\alpha_G(M_U)}-\frac{a}{2\pi} \ln \frac{\mu}{M_U}\label{eq3}\ee\\

\par\noindent
The  $\beta$-function coefficient
in eq.(3)  consists of gauge, matter and Higgs contributions,\\
\be a=a_{gauge}+a_{matter}+a_{Higgs}\label{eq4}\ee\\
 The gauge bosons of SO(10) in the adjoint representation $\underline{45}$,  
three generations of matter in the spinorial representations  
$\underline{16}$, and their superpartners contribute  as\\

\be a_{guage}=-24,~~a_{matter}=6\label{eq5}\ee\\
The Higgs contributions  of different  SO(10) irreducible representations 
are shown in Table 1.\\

\begin{table} 
\caption{ Contribution of Higgs representations to SUSY SO(10)  
$\beta$-function coefficient  for the GUT gauge coupling 
evolution}\label{tab1}
\begin{tabular}{llll}
\hline
Rep. & $a_{Higgs}$& Rep.&$a_{Higgs}$\\
\hline
$\un{10}$&1&$\un{45}\op\un{16}\op\un{\overline {16}}\op\un{10}$&13\\
$\un{54}$&12&$\un{54}\op\un{45}\op\un{16}\op\un{\ov{16}}\op\un{10}$&25\\
$\un{120}$&28&$\un{210}\op\un{16}\op\un{\ov {16}}\op\un{10}$&61\\
$\un{16}$&2&$\un{45}\op\rrep\op\un{10}$&79\\
$\un{45}$&8&$\un{210}\op\rrep\op\un{10}$&127\\
$\un{126}$&35&$\un{54}\op\un{45}\op\rrep\op\un{10}$&91\\
$\un{210}$&56&$\un{210}\op\un{54}\op\rrep\op\un{10}$&139\\
\hline
\end{tabular}
\end {table} 
\vspace{.5cm}
Noting that $a_{guage}+a_{matter}=-18$, use of eq.(5) in eqs.(3)-(4) gives at 
 $\mu=\Lambda > M_U$,\\

\be \frac{1}{\alpha_G(\Lambda)}=\frac{1}{\alpha_G (M_U)}+\frac{18}{2\pi}\ln\frac{\Lambda}{M_U}-\frac{a_{Higgs}}{2\pi}\ln\frac{\Lambda}{M_U}\label{eq6}\ee\\

If the gauge coupling constant encounters a Landau pole at $\Lambda$, $\alpha_G(\Lambda)\longrightarrow\infty $ and eq.(6) leads to \\

\be a_{Higgs}\le 18+ \frac{2\pi}{\ln(\frac{\Lambda}{M_U})}\times \frac{1}{\alpha_G(M_U)}\label{eq7}\ee\\

On the other hand the perturbative condition \\

\be \alpha_G(\Lambda)\le 1\label{eq8}\ee\\

leads to the constraint\\

\be a_{Higgs}\le 18+\frac{2\pi}{\ln (\frac{\Lambda}{M_U})}\left[\frac{1}{\alpha_G(M_U)}-1\right]\label{eq9}\ee \\

In the single step breakings of all SUSY GUTs\\
$M_U\simeq 2\times 10^{16}$ GeV, 
 ~${\alpha_G(M_U)}^{-1}\simeq 25$, 
 and the upper bound defined
by inequality (9)   has been  estimated \cite{22}.\\

For SUSY SO(10) with ~$\un{45}\op\un{16}\op\un{\ov {16}}\op\un{10}$, 
~$a_{Higgs}=13$ and the perturbative constraint remains valid for higher scales and perturbative grand unification is guaranted at least up to the Planck scale \cite{22}. However, for minimal 
SO(10) with $\un{210}
\op \rrep \op \un{10}$, $a_{Higgs}=127$ and the perturbation theory can not be guaranted to hold up to the Planck scale
in the grand desert model. Thus, in the single step breaking of SUSY
 SO(10) to MSSM,  whenever larger Higgs representations like $\rrep$ are used
to break the $SU(2)_R\times U(1)_{B-L}\subset$ SO(10) or
$SU(2)_R\times SU(4)_C\subset$ SO(10), leading to the seesaw mechanism and Majorana neutrino masses, the large contribution to the Dynkin indices violates perturbation theory at $\Lambda = {\rm few} \times 2\times 10^{16}$ GeV. This has led 
to the investigations of perturbative grand unification of SO(10) through the use of Higgs representations $\un {16}\op\un{\ov{16}}$ instead of $\rrep$  
in the supergrand-desert scenario \cite{22}. 

It is clear that in R-parity conserving SUSY SO(10) the Higgs contribution to
the $\beta$-function coefficient for the gauge coupling evolution satisfies
$a_{Higgs} > 71$. Noting that $\a_G(M_U) \simeq 0.043$ and demanding that 
perturbative condition is satisfied up to $\Lambda = M_{Pl} = 2\times 10^{18}$ GeV,
then the inequality (9) gives the lower bound,

\be M_U > 10^{17} GeV \nonumber \ee      

This lower bound on the unification scale has to be satisfied in any R-parity conserving SUSY SO(10) if the GUT gauge coupling is to remain perturbative up to the Planck scale. It is interesting to note that this lower 
bound  accidentally  matches the Higgsino mass limit obtained from the current 
experimental limit of the proton lifetime for 
$p \rightarrow K^+\overline{\nu}_{\mu,\tau}$.
    
In the four specific examples of Higgs representations shown in Table.1 
which correspond to R-parity conservation, the
Higgs contributions to the $\beta$-function coefficients in the respective 
cases and the inequality (9) give different values of lower bounds on the unification scale. In particular for the choices of the Higgs representations (I).       
 $\un{210}\op\rrep\op\un{10}$, (II). $\un{54}\op\un{45}\op\rrep\op\un{10}$,
(III). $\un{210}\op\un{54}\op\rrep\op\un{10}$, and (IV). 
$\un{45}\op\rrep\op\un{10}$, the lower bounds on the unification scale
turn out to be $\U = 5.8\times 10^{17}$ GeV, $\U = 3\times 10^{17}$ GeV,
$\U = 6.25\times 10^{17}$ GeV, and $\U = 1.5\times 10^{17}$ GeV, respectively.
Thus the smallest lower bound corresponds to the one for the Higgs 
representation $\un{45}\op\rrep\op\un{10}$ as expected with minimal 
contribution $a_{Higgs} = 79$. These lower bounds suggest that if the perturbative criteria on the GUT gauge coupling is to be satisfied, the unification scale has to be elevated by at least one order compared to the conventional value.
Further, the perturbative constraint has the implication that, in  R-conserving SUSY SO(10), the larger is the Higgs contribution to the $\beta$-function coefficient, the greater must be the unification scale. The lower bounds are to 
be satisfied irrespective of the SO(10) breaking to MSSM through a single
 step or through an intermediate gauge symmetry.     

In the next section  we ~show how the presence of 
$SU(2)_L\times SU(2)_R\times U(1)_{B-L}\times SU(3)_C$ 
intermediate gauge symmetry at higher scales yields perturbative 
SO(10) up to the
Planck scale  even if we use the Higgs representations 
$\rrep$ with or without 
$\un {210}$ or other Higgs representations such as $\un {54}$ and 
$\un {45}$
for high-scale breaking of SUSY SO(10).

\section{Threshold and gravitational corrections on mass scales}\label{sec3}

~~~It is clear from eq.(9) that, if in a specific GUT scenario the unification 
scale $\U $ can be closer to the Planck or the compactification scale than in the single step breaking case, the contribution of the Higgs representation 
to the RHS of (9) can be larger without violating the inequality. In 
\cite{20} the intermediate $G_{2213}$
~breaking in SUSY SO(10) was investigated,
 
\be &&{\rm SO(10)}\t {\rm SUSY}
\mathop{\longrightarrow}^{\Phi_U}_{\U}G_{2213}\t 
{\rm SUSY}\nonumber\\ &&\mathop{\longrightarrow}^{\rrep}_{\I}G_{213}\t 
{\rm SUSY}
\mathop{\longrightarrow}^{\un{10}}_{\Z}U(1)_{em}\t SU(3)_C\label{eq10}\ee
where the Higgs representations responsible for the GUT symmetry breaking
were chosen as $\Phi_U \equiv \un {210}$, 
or $\un {54}\oplus \un {45}$ which also 
break D-Parity at the GUT scale while permitting the left-right asymmetric
gauge group $G_{2213}$($g_{2L}\neq g_{2R}$) to survive down to the intermediate~scale ~\cite{23}. In such  an R-parity conserving symmetry breaking 
chain quite
significant threshold corrections  arising out of spreading of masses 
 around the intermedite scale and the GUT scale and gravitational corrections arising out of $5-{\rm dim}.$ operators induced by the Planck or the compactification scales \cite{24,25,26,27} were noted. In this section we estimate these effects in detail 
to explore the possibility of increasing $M_U$ which is necessary for the 
existence of
perturbative gauge coupling at higher scales. While the gravitational corrections originating from the 5-dim. operator due to $\un {210}$ was investigated in 
\cite{20}, in this work we investigate the corresponding effects due to
$\un {54}$ and  $\un {210}\op \un {54}$ while studying the threshold effects of the latter.
The evolution of gauge couplings in the two different mass ranges is expressed
as,    
\be{1\over\a_i(\Z)}&=&{1\over\a_i(\I)}+{a_i\over 2\pi}\ln{\I\over\Z}+\T_i-\D_i,\nonumber\\ 
i&=&1Y, 2L, 3C,  \label{eq11}\\ 
{1\over\a_i(\I)}&=&{1\over\a_i(\U)}+{a'_i\over 2\pi}\ln{\U\over\I}+\T'_i-\D'_i-\D_i^{(gr)},
\nonumber\\ i&=&2L, 2R, BL, 3C.\label{eq12}\ee 
where the second, third, and the fourth terms in the RHS of eqs.(11)-(12)
represent one-loop, two-loop, threshold, and gravitational corrections,
respectively \cite{20}. 
In eqs.(11)-(12) $a_Y = 33/5, a_{2L} = 1, a_{3C} = a'_{3C} = -3, a'_{2L} = 1, 
a'_{2R} = 5$ and  $a'_{BL} = 15$.   
The two-loop coefficients($b_{ij}$) below the intermediate scale and
($b'_{ij}$) above the intermediate scale have been obtained in 
\cite{20}. Below $\I$ the presence of $G_{213}$ in MSSM gives,

\be b_{ij}=\left(\ba{ccc} {199\over 25}&{27\over 5}&{88\over 5}\\ {9\over 5}&25&24\\ 
{11\over 5}&9&14\ea\right),\, i,j=1Y, 2L, 3C.\label{eq13}\ee
Above the intermediate scale, the two-loop beta-function 
coefficients in the presence of SUSY $G_{2213}$ symmetry are
 \be b'_{ij}&=&\left(\ba{cccc} 25&3&3&24\\ 
3&73&27&24\\ 9&81&61&8\\9&9&1&14\ea\right).\nonumber\\  
i, j&=&2L, 2R, BL, 3C.\label{eq14}\ee
These coefficients occur in  two-loop contributions represented by
$\T_i$ and $\T'_i$ in the two mass ranges,
\be\T_i&=&{1\over 4\pi}\sum_jB_{ij}\ln{\a_j(\I)\over\a_j(\Z)},\nonumber\\
\T'_i&=&{1\over 4\pi}\sum_jB'_{ij}\ln{\a_j(\U)\over\a_j(\I)},\nonumber\\
B_{ij}&=&{b_{ij}\over a_j},\hspace{.5cm} B'_{ij}={b'_{ij}\over a'_j}.\label{eq15}\ee
For the sake of simplicity we have neglected the Yukawa contributions to two-loop effects on gauge couplings.
While the functions $\D_i$ include threshold effects at $\Z$ and $\I$ with 
$$\D_i=\D^{(Z)}_i+\D^{(I)}_i$$ $\D'_i$ include threshold effects at $\U$.

It may be recalled that although  in nonsupersymmetric gauge theories 
threshold effects contain both constant terms as well as logarithmic terms,
it was noted in \cite{32} that the constant tems are absent in 
supersymmetric threshold corrections.  

In the presence of $G_{2213}$ intermediate symmetry the particle spectra of Higgs scalars, fermions, gauge bosons, and their 
superpartners with masses lighter than $\U$ are the same in all four cases being considered in this paper. Then, under the  assumption that all superheavy particles  with masses larger than $\U$ decouple from the Lagrangian, 
the contributions to the renormalization group evolutions of gauge and Yukawa  couplings  up to two loops below  $\U$ are identical in all the four cases,  
Case (I): $\un{210}\op\rrep\op\un {10}$, Case (II): 
$\un{54}\op\un {45}\op \rrep\op\un {10}$, Case (III): 
$\un {210}\op\un {54} \op \rrep \op\un {10}$, 
and Case (IV): $\un {45}\op \rrep\op\un {10}$. However, the GUT threshold 
and gravitational effects expressed through  $\D'_i$ and
$\D^{(gr)}_i$, respectively, differ from one choice of representation to 
another. 
\subsection{Threshold effects with effective mass parameters}\label{sec3.1}
We follow the method of effective mass parameters due to Carena, Pokorski, 
and Wagner \cite{28} to estimate threshold effects which  
have been also utilised to study such effects in SUSY SU(5) by introducing
two sets of  effective mass parameters , one set for the SUSY threshold and 
the other set  for the GUT threshold \cite{29}. In \cite{20} their effects 
have been examined on 
SUSY SO(10) with $G_{2213}$ intermediate symmetry by defining 
one set of effective mass parameters for each threshold. Although these
 parameters at the weak-scale SUSY threshold have been approximately 
estimated \cite{28,29}, no such estimations are available for higher thresholds and they would be assumed to deviate at most by a factor $6$($1/6$) from the 
corresponding scales.
Following the standard procedure, the effective mass parameters are defined 
through the following relations,    
\be \D^Z_i&=&\sum_{\a} {b^{\a}_i\over 2\pi}\ln{M_{\a}\over \Z}
={b_i\over 2\pi}\ln{M_i\over \Z},\nonumber\\
i&=&1Y,2L,3C;\, \mu=\Z;\label{eq16}\\
\D^I_i&=&\sum_{\a} {b'^{\a}_i\over 2\pi}\ln{M'_{\a}\over \I}
={b'_i\over 2\pi}\ln{M'_i\over \I},\nonumber\\ 
i&=&1Y,2L,3C;\, \mu=\I;\label{eq17}\\
\D'&=&\D^U_i=\sum_{\a} {b''^{\a}_i\over 2\pi}\ln{M''_{\a}\over \U}\nonumber\\
&=&{b''_i\over 2\pi}\ln{M''_i\over \U},\nonumber\\ 
i&=&2L,2R,BL,3C;\, \mu=\U;\label{eq18}\ee
where $\a$ refers to the actual $G_{213}$ submultiplet near $\mu = \Z$, $\I$ or
$G_{2213}$ submultiplet near $\mu = \U$ and $M_{\a}, M'_{\a}$ or $M''_{\a}$
refer to the actual component masses.
The three sets of effective mass parameters are $M_{\rm i}$, $M'_{\rm i}$, and
$M''_{\rm i}$. 
The coefficients $b'_i=\sum b'^{\a}_i$ 
and $b''_i=\sum b''^{\a}_i$ have been defined in eqs.(16)-(18) following 
 \cite{20,28}. The numbers $b^{\a}_i$ and $b'^{\a}_i$ refer to the 
contributions  
of the multiplet $\a$ to the $\beta$-functions of 
 $U(1)_Y$, $SU(2)_L$,  and   $SU(3)_C$ gauge couplings. Similarly $b''^{\a}_i$ 
refers to the contributions  
of the multiplet $\a$ to the $\beta$-functions of
$U(1)_Y$, $SU(2)_L$,  $SU(2)_R$,  $SU(3)_C$, and  $U(1)_{B-L}$ gauge
couplings \cite{20}.

The threshold effects on the mass scales $\I$ and $\U$ are then expressed in the form

\be \D\ln{\I\over \Z}&=& a\ln{M''_{2R}\over\U}
+b\ln{M''_{BL}\over\U}+c\ln{M''_{2L}\over\U}\nonumber\\
&&+d\ln{M''_{3C}\over \U}+e\ln{M'_{1Y}\over\I}-1.56,\nonumber\\
\D\ln{\U\over \Z}&=&a'\ln{M''_{2L}\over\U}
+b'\ln{M''_{3C}\over\U}+0.105\label{eq19}\ee
where the numerical values are due to the weak-scale SUSY threshold effects. 
The values of the parameters computed for the four different 
cases are,\\

Case(I): $\un{210}\op\rrep\op\un{10}$\\
\be (a,b,c,d,e)&=&(-25, -{57}/4, 130, -{355}/4, -9/4),\nonumber\\
(a', b')&=&(26, {-213}/8)\label{eq20}\ee\\
Case(II): $\un {54}\op \un {45}\op \rrep\op \un {10}$\\
\be (a,b,c,d,e)&=&(-{77}/4, -{45}/4, {405}/4, -{135}/2, -9/4),\nonumber\\
(a',b')&=&({81}/4, {-81}/4)\label{eq21}\ee \\
Case(III): $\un {210}\oplus \un {54}\op\rrep\op \un {10}$\\ 
\be(a,b,c,d,e)&=&(-{109}/4, -{61}/4, {565}/4, -95, -9/4),\nonumber\\
(a', b')&=&({113}/4, -{57}/2).\label{eq22}\ee\\
Case(IV): $\un {45}\op\rrep\op \un {10}$\\ 
\be(a,b,c,d,e)&=&(-{35}/4, -{31}/6, {185}/4, -{385}/{12}, -9/4),\nonumber\\
(a', b')&=&({37}/4, -{77}/8).\label{eq23}\ee\
Although  Cases (I)-(II) were derived in Ref.\cite{20} some numerical and 
typographical errors  have been corrected here while 
Cases (III)-(IV) are  new. 

\subsection{Gravitational corrections from dim.5 operators}\label{sec3.2}
 
  In this subsection we derive gravitational corrections in  Case(II) and 
Case(III) while such corrections in Case(I) were
discussed in  \cite{20}.
In addition to the renormalizable part of the Lagrangian of SUSY GUT,
a  5-dim operator can be induced either in  $4-{\rm dim}.$ gravity at the 
Planck scale
($M_C=M_{Pl}=2\times 10^{18}$ GeV)
or due to compactification of extra dimension(s) at scales 
$M_C=M_{CS}\sim 10^{17}$ GeV \cite{25}. 
\be {\cal {L}}_{gr}=-{\eta\over 2M_C}Tr\left(F_{\mu\nu}\Sigma F^{\mu\nu}\right)
\label{eq24}\ee
where, for example, $\Sigma \equiv {\un {210}}, {\un {54}} \subset 
{\rm SO(10)}$ 
that contribute to the GUT symmetry breaking near  $M_U$ and $M_C=$ 
compactification scale($M_{CS}$) of extra dimension(s), or the Planck 
scale($M_{Pl}$) in $4-{\rm dim}$. 
gauge theory. When   $\Sigma \equiv {\un {45}} \subset {\rm SO(10)}$ the 
contribution of the $5-{\rm dim}$. operator in eq.(24) identically vanishes.
We will confine to the Cases (I)-(III) for gravitational corrections.

Although there are no exact theoretical constraint on $\eta$
it could be positive or negative with plausible values up to $|\eta|\approx
{\rm O(10)}$.
Whereas $\un{210}$ and $\un{54}$ are present in cases I and II, respectively, 
both are present in case III. In 
 \cite{20}  gravitational effects were derived only for the Case (I)
corresponding to $\Sigma\equiv\un {210}$ with  a normalization factor $1/8$ instead of $1/2$ as given in eq.(24) \cite{26}. In order to compare with gravitational corrections resulting from eq.(24) with  $\Sigma\equiv\un {54}$ we 
 evaluate them for the Case (I) with the common normalization factor of $1/2$. 
In a number of earlier investigations the effects of such  operators on GUT predictions have been found to be quite significant \cite{18,20,25,26,27}.
In the presence of ${\rm SO(10)}\to G_{2213}$ such operators modify the GUT boundary condition on the coupling constants which has the general form at $\mu=\U$ ,\\

\be&&\a_{2L}(\U)(1+\ep_{2L})=\a_{2R}(\U)(1+\ep_{2R})\nonumber\\
&&=\a_{BL}(\U)(1+\ep_{BL})=\a_{3C}(\U)(1+\ep_{3C})\nonumber\\
&&=\a_G(M_U)\label{eq25}\ee 
These boundary conditions lead to the corresponding gravitational corrections
on the four gauge couplings, 

\b\D_i^{gr}=-{{\ep}_i\over\a_G},\, i=2L, 2R, BL, 3C\label{eq26}\e
Then using the procedure
of \cite{20}, analytic formulas for the gravitational corrections of the 
two mass scales are derived,

\be\left(\ln{\I\over \Z}\right)_{gr}&=&{2\pi(A'\ep'-A\ep'')\over\a_G(AB'-A'B)},
\nonumber\\
\left(\ln{\U\over \Z}\right)_{gr}&=&{2\pi(B\ep''-B'\ep')\over\a_G(AB'-A'B)}.
\label{eq27}\ee
where

\be B&=&B'={5\over 3}a_Y-{2\over 3}a'_{BL}-a'_{2R},\nonumber\\
A&=&a'_{2R}+{2\over 3}a'_{BL}-{5\over 3}a'_{2L},\nonumber\\
A'&=&a'_{2R}+{2\over 3}a'_{BL}+a'_{2L}-{8\over 3}a'_{3C},\nonumber\\
{\ep}''&=&{\ep}_{2L}+{\ep}_{2R}+{2\over 3}{\ep}_{BL}-{8\over 3}{\ep}_{3C},\nonumber\\
{\ep}'&=&{\ep}_{2R}+{2\over 3}{\ep}_{BL}-{5\over 3}{\ep}_{2L}\label{eq28}\ee
We will need the numerical values of $A, A', B, B'$ defined through eq.(28)
which are the same in all R-parity conserving cases with $G_{2213}$
intermediate gauge symmetry,

\be A &=& 40/3,~~A' = 24, \nonumber \\
B &=& B' = -4 \label{eq29}\ee
With the generalized formulas given by  eqs.(25)-(28) and
the numerical values given in eq.(29) we 
discuss specific gravitational corrections in  three different cases
as given below.\\

\par\noindent Case (I): $\un{210} \op \rrep \op\un {10}$\\
 
In this case $\Sigma\equiv\un {210}$ and we denote the unknown parameter in eq.(24)
as $\eta=\eta_1$. After taking into account a factor $4$
in the normalization of the gauge kinetic term \cite{26,27} and using an
approximate relation between the GUT-scale VEV $\phi_0$ and the degenerate 
masses of superheavy gauge bosons, $M_U \approx {(2/9)}^{1/2}g_G\phi_0$, 
we have\\ 

\be \epsilon_{2R}&=& -\epsilon_{2L} = -\epsilon_{3C} = \frac{1}{2}\epsilon_{BL} = \epsilon_1,\nonumber\\
\epsilon'&=& \epsilon_{2R}+\frac{2}{3}\epsilon_{BL}-\frac{5}{3}\epsilon_{2L}=
4\ep_1,\nonumber\\
\epsilon'' &=& \epsilon_{2R}+\epsilon_{2R}+\frac{2}{3}\epsilon_{BL}-\frac{8}{3}\epsilon_{3C}=4\ep_1\label{eq30}\ee\\
where
\be\ep_1 = \frac{3\eta_1}{4}\frac{M_U}{M_C}\frac{1}{\sqrt{4\pi \alpha_G}}\label{eq31}\ee\\
Using  eqs.(29)-(30) in eq.(27) gives

\be(\ln\frac{M_I}{M_Z})_{gr}&=&\frac{2\pi\epsilon_1}{\alpha_G},\nonumber\\    
(\ln\frac{M_U}{M_Z})_{gr}&=&0\label{eq32}\ee\\    
which were derived in  \cite{20} but with a different normalization factor for $\ep_1$.\\

\par\noindent Case (II): $\un {54}\op \un{45} \op \rrep \op\un {10}$\\

In this case $\un {45}\subset$ SO(10) does not contribute to the ${\rm dim}.5$ 
operator of eq.(24). Using $\Sigma\equiv\un {54}$ and denoting $\eta= \eta_2$ in eq.(24), we derive

\be\ep_{3C}&=& \ep_{BL} = \ep_2,\nonumber\\
\ep_{2L}&=& \ep_{2R} = -\frac{3}{2}\ep_2,\nonumber\\
\ep'&=&\ep_{2R}+\frac{2}{3}\ep_{BL}-\frac{5}{3}\ep_{2L}=\frac{5}{3}\ep_2,\nonumber\\
\ep''&=& \ep_{2L}+\ep_{2R}+\frac{2}{3}\ep_{BL}-\frac{8}{3}\ep_{3C}=-5\ep_2\label{eq33}\ee
where    
\be\ep_2 = \frac{3\eta_2}{4}\frac{M_U}{M_C}\frac{1}{\sqrt{15\pi\alpha_G}}\label{eq34}\ee
Using eq.(29)  and eqs.(33)-(34) in 
eq.(27), we get,\\    

\be(\ln\frac{M_I}{M_Z})_{gr}&=& \frac{5\pi\ep_2}{\a_G},\nonumber\\    
(\ln\frac{M_U}{M_Z})_{gr}&=& \frac{5\pi\ep_2}{4\a_G}\label{eq35}\ee   
Eq.(35) has the implication  that if we attempt to change the
unification mass by one order purely by gravitational corrections, then the intermediate scale would change by approximately four orders.\\

\par\noindent Case (III): $\un{210} \op\un {54} \op \rrep \op\un {10}$\\

The  importance of this case emphasizing the presence of $\un {54}$ in addition to $\un {210}$ for realistic SUSY SO(10) breaking leading to Type II seesaw dominance for neutrino masses has been elucidated  in  \cite{21} 
in the single step breaking case. 
In our case with $G_{2213}$ intermediate symmetry   both $\un{54}$ and 
$\un{210}$ contribute separately to the 
 ${\rm dim}.5$ operator with \\
\be{\cal {L}}_{gr} &=&-\frac{\eta_1}{2M_C}Tr(F_{\mu\nu}\phi_{210}F^{\mu\nu})\nonumber\\
&&-\frac{\eta_2}{2M_C}Tr(F_{\mu\nu}\phi_{54}F^{\mu\nu})\label{eq36}\ee\\
Then 

$$\D_i^{gr}=-(\ep_i^{54}+\ep_i^{210})/\a_G$$, \nonumber
$$i=2L, 2R, BL, 3C$$\nonumber\\
The relations (32) and (35) hold separately leading to  

\be(\ln\frac{M_U}{M_Z})_{gr} &=& \frac{5\pi\epsilon_2}{4\alpha_G},\nonumber\\
(\ln\frac{M_I}{M_Z})_{gr} &=& \frac{5\pi\epsilon_2}{\alpha_G} + \frac{2\pi\epsilon_1}{\alpha_G}.\label{eq37}\ee\\
Comparing eqs.(31) and (34) gives $\ep_1/\ep_2 =(15/4)^{1/2}\eta_1/\eta_2$.  
In the next section we use these  results to study the
effects of gravitational corrections on SO(10) gauge coupling.
 
\section{Perturbative SO(10) gauge coupling at higher scales}\label{sec4}
 
In all the three cases the same lighter components  contained in 
$\rrep\op \un {10}$ contribute to the one-loop and two-loop $\beta$-function 
coefficients below the GUT scale and none of the components in ${\un {210}}, 
{\un {54}}$, or ${\un {45}}$ contribute to 
large runnings of the gauge couplings. Thus, ignoring threshold and gravitational corrections, the two-loop solution of RGEs  is the same for all the four cases with\\

\be M_I^0 &=& 10^{15.2} ~GeV,  M_U^0 = 10^{16.11} ~GeV, \nonumber \\
\a_G^0 &=& 0.043 \label{eq38}\ee\\
Then adding  threshold and gravitational corrections to two-loop solutions
the mass scales are expressed  as\\

\be \ln\frac{M_U}{M_Z}&=&\ln\frac{M_U^0}{M_Z}+\D\ln\frac{M_U}{M_Z}+(\ln\frac{M_U}{M_Z})_{gr},\nonumber\\ 
\ln\frac{M_I}{M_Z}&=&\ln\frac{M_I^0}{M_Z}+\D\ln\frac{M_I}{M_Z}+(\ln\frac{M_I}{M_Z})_{gr}.\label{eq39}\ee  
     
When we include corrections mentioned in Sec.3 through eq.(39), the resulting mass scales are modified in each case. The incresed value of $M_U$  
then  extends the range of perturbative SO(10) gauge coupling up to the 
Planck scale. In what follows we discuss some examples of such solutions  
in each case. 

The mass scales obtained including threshold corrections are 
denoted as $M^{(1)}_i$ and those otained  including both the threshold and gravitational corrections are denoted as $M^{(2)}_i$(i=I, U) \\
\par\noindent
Case (I): As shown in Sec.2 the lower bound on the unification mass in this 
case is $5.8\times 10^{17}$ GeV. Using threshold and gravitational corrections we examine how far this constraint can be satisfied.   
Using the effective mass parameters

\be M''_{2L}&=&M_U, ~M''_{3C}=0.87M_U, \nonumber \\
M''_{2R} &=& 1.5M_U, M''_{BL} = 1.8M_U, \nonumber \\
M'_{1Y} &=& M_I \nonumber \ee
we obtain 
including only  threshold effects,

$$\U^{(1)}=6.54\times 10^{17} GeV,\hspace{.5cm}\I^{(1)}=7\times 10^{11} GeV$$    
Using this modified value, $\U=\U^{(1)}=6.54\times 10^{17}$ GeV
eq.(6) gives the perturbative value of the GUT-gauge coupling at 
$\Lambda = M_{Pl}$  with ~~$\a_G(M_{Pl})=0.587$.
 The effects are more(less) prominent if the mass gap of the effective mass parameters are increased(decreased) for which the values of the corresponding 
gauge coupling will be smaller(larger). It is easily checked that the 
inequality (9) is  satisfied. 
  Since the gravitational corrections do not affect the GUT scale, but affect only the intermediate scale which is of the same order as the right-handed 
neutrino mass, in this case any desired value of the intermediate scale 
matching the scale of leptogenesis, or the Pecei-Quinn symmetry breaking scale,
 or even a value close to the  minimal GUT scale can be obtained. Thus the 
 model is potentially  interesting  from the point of view of  neutrino 
physics, leptogenesis and 
 strong CP-violation. ~Other examples of solutions for this case are shown in 
Table.2.\\

\par\noindent
Case (II): As shown in Sec.2 the value of $a_{Higgs} = 91$ in this case gives the lower bound $\U > 3\times 10^{17}$ GeV.
To examine how far threshold and gravitational corrections may allow such high unification scales, at first we consider only threshold corrections. Using the 
effective mass  parameters\\ 
\be M''_{2L}&=& M_U, M''_{2R}=1.7M_U, M''_{BL}= 2M_U,\nonumber\\
M''_{3C}&=&0.87M_U,  M'_{1Y}=M_I\nonumber\ee
gives including  threshold corrections but ignoring gravitational corrections,\\

$$\I^{(1)}=M_R=2.91\times 10^{11}GeV, \hspace{.5cm}\U^{(1)}=3.85\times 10^{17} 
GeV$$\\
Clearly the mass gaps near the GUT scale are reasonably small and are confined
between $0.87\U$ and $2\U$. Now adding gravitational corrections with
$\eta_2=3.0$ gives,
$$\I^{(2)}=9.31\times 10^{12} ~GeV,~~\U^{(2)}=8.95\times 10^{17} GeV \nonumber
 $$
Using the value of $\U= \U^{(2)}=8.95\times 10^{17}$ GeV we obtain from  
eq.(6) the perturbative value of the gauge coupling ,
$\a_G(M_{Pl}) =  0.084$. 
Evaluating the  RHS of inequality (9) gives \\
$$a_{Higgs}< 180 \nonumber $$\\ 
 Noting from Table.1 that for this case $a_{Higgs}=91$ it is clear that 
 inequality (9) is satisfied ensuring  perturbativity of 
SO(10) gauge coupling up to the Planck scale. Another example of solution 
including gravitational correction is given in Table.2\\
\par\noindent
Case (III):
As shown in Sec.2  for this case 
$a_{Higgs}=139$ and  (9) gives the lower bound  
$M_U \ge 6.25 \times 10^{17}$ GeV to ensure perturbative gauge coupling up to
the Planck scale. The necessity of both $\un {210}$ and  $\un {54}$ for 
realistic SUSY $SO(10)$ breaking directly to MSSM has been emphasized in  
\cite{21}. With $G_{2213}$ intermediate breaking   
this case appears to be  interesting as it shows the possibility
that dominant gravitational corrections with marginal or negligible threshold 
effects
can elevate  the GUT scale closer to  the Planck 
scale \cite{25}. 
Although, in principle, threshold effects  are  somewhat larger in this case 
compared to 
the Cases (I)-(II) and (IV) because of the presence of extended size of Higgs representations, their actual values are controlled by the choice of the mass gap in the effective mass parameters. For example, using the effective mass parameters,\\
\be M''_{2L}&=&M_U, ~M''_{3C}=0.87M_U,\nonumber\\ 
~M''_{2R}&=&1.6M_U, M''_{BL}=1.6M_U,\nonumber\\
~M'_{1Y}&=&M_I \nonumber \ee
we obtain including only threshold effects,
$$\U^{(1)}=7.57\times 10^{17} ~GeV,\hspace{.2cm}\I^{(1)}=3.92\times 10^{11} 
~GeV$$    
Then eq.(6) gives the perturbative gauge coupling\\ 
$\a_G(M_{Pl}) \simeq 0.25$.\\
Further addition of gravitational corrections with $\eta_1=-3.0$ and 
$\eta_2= 5.0$ gives higher values of the unification scale  closer to $M_{Pl}$,
$$\U^{(2)} = 2.95\times 10^{18}~GeV, \hspace{.2cm} \I^{(2)} =6.96\times 10^{12} ~GeV, \nonumber $$
Using  this high  value of the unification  scale  
$\U=\U^{(2)}=2.95\times 10^{18}$
GeV we obtain from eq.(6) the perturbative value of the gauge coupling  
$\a_G(M_{Pl})\simeq .049$. We also note that the perturbative inequality (9)
is easily satisfied with $\Lambda = M_{Pl}$ . 
Another example of such solution for this case is shown in Table.2 where both threshold and gravitational corrections have been included.

Now we show that with negligible GUT threshold corrections
but with the inclusion of gravitational corrections alone in this case 
it is also possible
 to obtain high values of the unification scale and perturbative gauge coupling up to the Planck scale. 
For the sake of simplicity  ignoring 
all high scale threshold corrections by choosing $M'_{1Y}=\I$ and $M''_i=\U$
(i=2L, 2R, BL, 3C) and  using $\eta_1=-20.0$ and $\eta_2=14.2$ leads to
$\ep_1=-0.128$ and $\ep_2=0.047$. Then eq.(36) gives 
 $(\ln\frac{M_U}{M_Z})_{gr}=4.33$ and $(\ln\frac{M_I}{M_Z})_{gr}
=-1.49$. When added to two-loop solutions including the  weak-scale SUSY 
threshold corrections  we obtain,
$$ M_U^{(2)}= 8.61 \times 10^{17} ~GeV, \hspace{.2cm} M_I^{(2)}=1.09\times 10^{14} ~GeV $$    
Using $\U=\U^{(2)}=8.61\times 10^{17}$ GeV in eq.(6) gives the perturbative
value of the gauge coupling at $\Lambda=M_{Pl}$ with $\a_G(M_{Pl})=0.175$.
  We find that the RHS of the  (9) is 
~$\simeq 190$ as compared to the value $a_{Higgs}=139$ for this case and
the perturbative inequality is satisfied. Thus, 
including gravitational corrections alone the $SO(10)$ model with such choice of Higgs representation 
guarantees perturbative SUSY $SO(10)$ gauge coupling up to the Planck scale.

\par\noindent Case (IV): As shown in Sec. 2, $a_{Higgs}=79$ 
through  (9) gives the lower bound $\U \ge 1.5\times 10^{17}$ GeV 
in this case to ensure
perturbative gauge coupling up to Planck scale. As there is no gravitational corrections due to the $5-{\rm dim}$. operator for this case we will consider
only threshold corrections. Using 

\be M''_{2L}&=& 1.5M_U, M''_{2R}= M''_{BL}= 3.5M_U,\nonumber\\
M''_{3C}&=&M_U,  M'_{1Y}=M_I\nonumber\ee
we obtain\\ 
$$\I^{(1)}=1.2\times 10^{15}GeV, \hspace{.5cm}\U^{(1)}=4.7\times 10^{17} 
GeV$$\\
Using $\U=\U^{(1)}=4.7\times 10^{17}$ GeV in eq. (6) gives the perturbative 
gauge coupling at the Planck scale with $\a_G(M_{Pl}) \simeq 0.10$.
The RHS of  (9) is found to be $\simeq 119$ and the inequality is
satisfied.

\section{\bf Proton lifetime predictions}\label{sec5}

As pointed out in Sec.1, the experimental lower limit on the proton lifetime 
 for the decay mode $p\to e^+\pi^0$ mediated by superheavy gauge bosons or 
equivalently through the effective ${\rm dim}.6$ operator sets a lower limit on the GUT scale,  $M_U \ge 5.6\times 10^{15}$ GeV which is easily satisfied 
in the supergrand desert scenario for which, excluding threshold or gravitational corrections, $M_U=2\times 10^{16}$ GeV. 
The lower bounds on $\U$ obtained in Sec.2 for the Cases (I)-(IV), purely from 
the requirement of perturbativity of the $SO(10)$ gauge coupling  up to the 
Planck scale, are found to 
be satified by the RG solutions for the mass scales when threshold corrections,
or gravitational corrections, or both are included in the intermediate scale 
models.  In the 
Case (IV) for which the Higgs representations $\un{45}\op\rrep\op\un{10}$
have the smallest size among all the four cases,  the solutions of RGEs for 
the mass scales are consistent with 
the lower bound $\U \ge 1.5\times 10^{17}$ GeV when threshold effects 
are included.
In each of the four cases  the corresponding lower bound on the unification scale translates into a lower bound on proton lifetime.  
 The shortest of these 
lower bounds on the proton lifetime occurs in the Case (IV),     

\be\tau \left(p\rightarrow e^+\pi^0\right) \ge 2.1 \times 10^{39} \rm {years}
\label{eq40}\ee
In the Cases (I)-(III) the lifetimes are longer than this value as 
can be approximately estimated using Table 2. These analyses suggest that 
the decay mode $p\rightarrow e^+\pi^0 $ which has lifetime at least $6$
orders longer than the current limit is inaccessible to 
experimental observation.

 Supersymmetric decay modes of the proton such as 
$p\rightarrow {\rm K}^+\bar{\nu_{\mu}}$, 
$p\rightarrow {\rm K}^+\bar{\nu_{\tau}}$ and others are    
characteristic predictions in SUSY GUTs \cite{30}. These decays 
are  mediated by Higgsinos 
$(T_{\tilde {C}})$ which are superpartners of colour triplet Higgs scalars $(T_C)$ having superheavy masses near the GUT scale. 
  As pointed out  the experimental lower limit on the proton lifetime given in 
eq.(2) sets the 
lower  bound on the superheavy colour triplet Higgsino mass,   
$M_{T_{\tilde {C}}} \ge 10^{17}$ GeV.

In SUSY SU(5) there is  one such pair of Higgssinos which are superpartners of
Higgs colour triplets contained in ${\un 5}\op\ov {\un 5} \subset$ SU(5); in 
SUSY SO(10) models the colour triplet Higgs may be treated as linear 
combination  of
the triplets cotained in $\un {10}$, $\rrep$ and  $\un {45}$, or $\un {54}$, 
or $\un {210}$ depending upon the choice of specific Higgs representations 
used to break the GUT symmetry to $G_{2213}$ \cite{19}.  
 For the sake of simplicity we ignore  finer details of calculations and
 give  plausibility arguments to show that for these decays 
 governed by the effective ${\rm dim}.5$ operators 
proton lifetimes ranging from 
the present experimental limit to several orders longer can be a natural 
prediction of the intermediate breaking scenario.

\par In a supergrand desert model like SUSY $SU(5)$,  the constraint on the colour triplet Higgsino mass is obtained using the unification condition
including  threshold corrections: $g_{\rm G}(\Lambda_U) = g_{1Y}(\Lambda_U) +
\Delta_{1Y}(\Lambda_U)= g_{2L}(\Lambda_U) + \Delta_{2L}(\Lambda_U)$
 where $g_G=$ GUT gauge coupling and $\Lambda_U = $ 
GUT scale. This leads to the constraint  
$g_G^{-2}(\Lambda_U)- g_{3C}^{-2}(\Lambda_U) 
=(3/{20\pi^2})\ln(M_{T_{\tilde C}}/{\Lambda_U})$ and 
$M_{T_{\tilde C }}\simeq$ few $\times
10^{15}$ GeV \cite{33}. However, including gravittional corrections large 
increase
of the Higgsino mass even up to four  orders of magnitude has been suggested 
in SUSY  SU(5) \cite{18}.

But in the presence of $G_{2213}$ intermediate symmetry in the mass range
$\mu = \I - \U$, the GUT scale  constraint equating $g_{1Y}$ and $g_{2L}$ is 
absent since the $U(1)_{Y}$-gauge coupling splits 
above the scale $\I$
into two separate unconstrained gauge couplings ,

\be \frac{1}{g_{1Y}^2(\mu)}=\frac{2}{5}\frac{1}{g_{BL}^2(\mu)}+\frac{3}{5}
\frac{1}{g_{2R}^2(\mu)}, ~\mu=\I-\U \nonumber \ee 
   
As the gauge symmetry near $\Lambda_U$ is no longer the SM, but
it is $G_{2213}$  the simple SU(5) relation among $g_G$, $g_{3C}$ and  
$M_{ T_{\tilde C}}$ is 
no longer valid. Further, unlike SU(5) where the Higgs colour triplet
and anti-triplet are confined to its Higgs representations, 
${\un 5}\op \ov {\un 5}$, in SO(10) their number is much more as they can originate from Higgs representations like $\un {10}$, $\rrep$, $\un {45}$, 
$\un {54}$, and $\un {210}$.  
In view of these there is no similar precision constriant 
on $M_{T_{\tilde C}}$ as in SUSY SU(5)
originating from
gauge coupling unification. In the presence of such two-step breaking 
through $G_{2213}$ intermediate gauge symmetry  the value of 
$M_{T_{\tilde C}}$  can easily exceed $10^{17}$ GeV.

Since our lower bounds needed for 
 perturbative gauge coupling up to the Planck scale as shown in Sec.4  are
in the range,

\be \U \ge (1.5-6.2)\times10^{17} \rm {GeV}\nonumber\ee 
and the lifetime for the supersymmetric decay modes are proportional to  $M_{T_{\tilde {C}}}^2$,   the lower bound on lifetimes are expected to be longer by 
factors 
ranging between $2.2$ and $38$  compared to the single-step breaking scenario.
This is  due to the natural expectation that without additional fine tuning  
all superheavy components including the colour triplets would have masses 
close   to $\U$.
Thus, the  criteria of perturbative gauge coupling up to
the Planck scale which are easily met by threshold or gravitational corrections
in the four cases of R-parity conserving SUSY SO(10), constrain  the unification scales  with  
$\U \ge (1.5-6.2)\times10^{17}$ GeV which in turn predict  for the supersymmetric decay modes of the proton,
\be\tau(p \rightarrow K^+\ov{\nu}_{\tau})\ge (2-9)\times 10^{34} \rm years
\label{eq41}\ee

But it is well known that even without additional fine tuning
 the superheavy components could be easily few times lighter or heavier than 
$\U$. Stretching this factor to the value of $\simeq 1/6$ or $6$ 
the lower limit on the proton lifetime has a wider range starting from the 
current experimental limit up  to a value which is 2-3 orders longer.  
 
 It is interesting to note that high-scale perturbative  renormalization group relations  (6) or (9) 
and the R-parity conservation in SUSY SO(10) predict these lower bounds on 
the unification scales, the smallest one being 
$\U \simeq 1.5\times10^{17}$ GeV.  The resulting  longer values of  proton litime predictions are 
 consequences of  generalized perturbative criteria in R-parity conserving SUSY SO(10) which are also solutions to perurbative renormalization group equations including threshold or gravitational corrections.

\section{\bf Summary and conclusion}\label{sec6}

 SUSY SO(10) with $\rrep$ and other Higgs representations in the case of  single-step breaking to MSSM has many attractive features for all fermion masses and mixings while ensuring R-parity conservation. But the 
popular argument raised against the model is that it violates perturbative  gauge theory as the GUT coupling blows off even at mass scales few times larger than the conventional GUT scale.
In this paper we have shown that the requirement that the GUT gauge coupling 
remains  perturbative  up to the Planck scale imposes lower bounds on the unification scale 
which are at least one order larger than the conventional GUT scale.
We have shown that the solutions to RGEs respecting these lower bounds are in fact possible if the  threshold and/or gravitational corrections are included.
 The four different  models discussed here ensure perturbative gauge coupling 
at least up to the Planck scale. The proton lifetime for 
$p\rightarrow e^+\pi^0$ becomes longer  at least by nealy $6$ orders 
of magnitude compared to
the current experimental limit. For the supersymmetric decay modes a wide 
range of lifetimes is possible extending from the current experimental 
limit  up to values 2-3 orders longer. These consequences follow 
without any additional fine tuning and by adopting the plausible 
criteria that is : in the presence of the intermediate gauge symmetry, 
all  superheavy masses including the colour triplet 

Higgsinos
have masses similar to the new high values of the unification scales. Although we have used the value of reduced Planck scale for this analysis, we have checked that our method also works 
with $M_{Pl} = 1.2\times 10^{19}$ GeV as defined by the Particle Data Group.
      
 Due to high values of the intermediate scale, the success of 
explanation of fermion masses and mixings are expected to be similar to 
single step breaking case, but the additional advantages of  high unification 
scale is that it ensures perturbative SUSY $SO(10)$ with R-parity conservation
at least  up to the Planck scale and 
increases the stability of the proton. A different scenario for the 
increase of   the proton stability in 
single step breaking  of
 SUSY SO(10) with R-parity conservation   has been 
suggested recently  by introducing specific 
textures \cite{19} where perturbative condition on SUSY $SO(10)$ gauge
coupling holds up to  $\mu= {\rm few}\times 2 \times 10^{16}$ GeV.\\

{\it Acknowledgements.} The work of M. K. P. is supported by the project No.
SP/S2/K-30/98 of Department of Science and Technology, Govt. of India. 

\begin{table*}
\caption{Perturbative SO(10) gauge coupling at higher scales including
threshold and gravitational corrections to two-loop solutions for which
$M^0_I=10^{15.20}$ GeV and $M^0_U=10^{16.11}$ GeV. The mass scales $M^{(1)}_i
(i=I,U)$
have been obtained including threshold corrections and $M^{(2)}_i(i=I,U)$ 
including both threshold and gravitational  corrections.}\label{tab2}
\begin{tabular}{ccccccccc}\hline\noalign{\smallskip}
Higgs Rep.&Mass parameters&$\I^{(1)}$(GeV)&$\U^{(1)}$(GeV)&$\eta_1$&$\eta_2$&$\I^{(2)}$(GeV)&$\U^{(2)}$(GeV)&$\a_G(M_{Pl})$\\ 
\hline\noalign{\smallskip}
\multicolumn{1}{c}{$\un{210}\op$}&$M'_{1Y}=\I, M''_{2L}=\U$&&&$0.0$&--&$7
\times 10^{11}$&$6.54\times 10^{17}$&\\\noalign{\smallskip}
$\rrep$&$M'_{2R}=1.5\U,M''_{BL}=1.8\U$&$7\times 
10^{11}$&$6.54\times 10^{17}$&$0.5$&--&$1.08\times 10^{12}$&$6.54\times 10^{17}$&$0.587$\\\noalign{\smallskip}
$\op\un{10}$&$M''_{3C}=0.87\U$&&&$5.0$&--&$5.49\times 10^{13}$&
$6.54\times 10{17}$&\\
\cline{2-9}
&&&&&&&&\\\noalign{\smallskip}
&$M'_{1Y}=2\I, M''_{2L}=2.5\U$&&&$1.5$&--&$1.05\times 10^{13}$&
$8.38\times 10^{17}$&$0.125$\\
\noalign{\smallskip}
&$M'_{2R}=4.5\U, M''_{BL}=3.76\U$&$2.84\times 10^{12}$&$8.38\times 
10^{17}$&$4.9$&--&$2.05\times 10^{14}$&$8.38\times 10^{17}$&\\
\noalign{\smallskip}
&$M''_{3C}=2.1\U$&&&&&&\\\noalign{\smallskip}
\hline\noalign{\smallskip}
&&&&&&&&\\\noalign{\smallskip}
$\un{54}\op\un{45}\op$&$M'_{1Y}=\I, M''_{2L}=\U$&&&$--$&$3.0$&$9.31\times 
10^{12}$&$8.95\times 10^{17}$&$0.084$\\\noalign{\smallskip}
$\rrep$&$M'_{2R}=1.7\U, M''_{BL}=2\U$&$2.91\times10^{11}$&$3.85
\times 10^{17}$&&&&&\\\noalign{\smallskip}
$\op\un{10}$&$M''_{3C}=\U$&&&$--$&$6.5$&$5.3\times 10^{14}$&$2.1\times 10^{18}$&$0.047$\\\noalign{\smallskip}
\hline\noalign{\smallskip}
&&&&&&&&\\\noalign{\smallskip}
$\un{210}\op\un {54}\op$&$M'_{1Y}=\I, M''_{2L}=\U$&&&$-1.5$&$2.5$&$1.65
\times 10^{12}$&$1.5\times 10^{18}$&$0.09$\\\noalign{\smallskip}
$\rrep$&$M'_{2R}=1.6\U,M''_{BL}=1.6\U$&$3.92\times 10^{11}$&$7.57\times 10^{17}
$&&&&&\\
\noalign{\smallskip}
$\op\un{10}$&$M''_{3C}=0.87\U$&&&$-3.0$&$5.0$&$6.96\times 10^{12}$&$2.95\times
10^{18}$&$0.049$\\
\cline{2-9}
&&&&&&&&\\\noalign{\smallskip}
&$M'_{1Y}=\I,M''_i=\U$&$3.3\times10^{14}$&$1.43\times10^{16}$&$-20.0$&$14.2$&
$1.09\times 10^{14}$&$8.61\times 10^{17}$&$0.175$\\\noalign{\smallskip}
&i=2L, 2R, BL, 3C&&&&&&&\\\noalign{\smallskip}
\hline
$\un{45}\op$&$M'_{1Y}=\I, M''_{2L}=1.5\U$&&&&&&&\\\noalign{\smallskip}
$\rrep$&$M'_{2R}=M''_{BL}=3.5\U$&$1.2\times 10^{15}$&$4.73\times 10^{17}
$&--&--&--&--&$0.10$\\
\noalign{\smallskip}
$\op\un{10}$&$M''_{3C}=\U$&&&&&&&\\\noalign{\smallskip}
\hline
\end{tabular}
\end{table*}

\end{document}